\documentclass[aps,prl,twocolumn,floatfix,showpacs]{revtex4}

\usepackage{graphics}
\usepackage{bm}

\def\PRA{{Phys.~Rev.~A} }

\def\JPB{{J.~Phys.~B} }
\def\PRL{{Phys.~Rev.~Lett.} }

\newcommand{\myscaleboxb}[1]{\scalebox{0.35}[0.25]{#1}}
\newcommand{\myscaleboxc}[1]{\scalebox{0.5}[0.35]{#1}}

\newcommand{\be}{\begin{equation}}
\newcommand{\bea}{\begin{eqnarray}}
\newcommand{\eea}{\end{eqnarray}}
\newcommand{\ee}{\end{equation}}

\begin{document}

\title{Extraction of the species dependent dipole moment from
high-order harmonic spectra in rare gas atoms }

\author{Anh-Thu Le,$^1$
Toru Morishita,$^{1,2}$ and C.~D. Lin$^1$}

\affiliation{$^1$Department of Physics, Cardwell Hall, Kansas
State University, Manhattan, KS 66506, USA\\
$^2$Department of Applied Physics and Chemistry, University of
Electro-Communications, 1-5-1 Chofu-ga-oka, Chofu-shi, Tokyo,
182-8585, Japan and
PRESTO, Japan Science and Technology Agency, Kawaguchi, Saitama 332-0012,
Japan}

\date{\today}

\begin{abstract}
Based on high-order harmonic generation (HHG) spectra obtained
from solving the time-dependent Schr\"odinger equation for atoms,
we established quantitatively  that the HHG yield can be expressed
as the product of a returning electron wave packet and the
photo-recombination cross sections, and the shape of the returning
wave packet is shown to be largely independent of the species. By
comparing the HHG spectra generated from different targets under
identical laser pulses, accurate structural information, including
the phase of the recombination amplitude, can be retrieved. This
result opens up the possibility of studying the target structure
of complex systems, including their time evolution, from the HHG
spectra generated by short laser pulses.

\end{abstract}


\maketitle

 When an atom is subjected to a strong driving laser field, one of
 the most important nonlinear response processes is the generation of high-order
 harmonics. In the past decade, high-order harmonic generation (HHG) has been used
 for the production of single attosecond pulses \cite{drescher,sekikawa,sansone}
 and attosecond pulse trains \cite{apt},
 thus opening up new opportunities for attosecond time-resolved
 spectroscopy. HHG is understood using the
 three-step model (TSM) \cite{corkum,kulander,lewenstein} -- first the electron is released by tunnel
 ionization; second, it is accelerated by the oscillating electric
 field of the laser and later driven back to the target
 ion; and third, the electron recombines with the ion to emit
 a high energy photon. A semiclassical formulation of the TSM
 based on the strong-field approximation (SFA) is given by
 Lewenstein {\it et al} \cite{lewenstein}.
 In this model (often called Lewenstein model),
 the liberated continuum electron experiences the full effect from the laser
 field, but not from the ion that it has left behind. In spite of
 this limitation, the SFA model has been used quite successfully,
 in particular, for analysis of the attosecond synchronization of high
 harmonics, see Mairesse {\it et al} \cite{mairesse} and references therein.
 However, since the continuum electron recombines when it is near the parent ion,
 the neglect of electron-ion interaction in the SFA model is rather
 questionable.

  According to the TSM, the last step of HHG
  is analogous to the radiative recombination process in
  electron-ion collisions. Thus one may write the HHG signal as
\begin{equation}
  S(\omega)=W(E)\times |d(\omega)|^2
\end{equation}
where $d(\omega)$ is the photo-recombination (PR) transition
dipole and $W(E)$ is the returning ``electron wave packet".
Electron energy $E$ is related to the emitted photon energy
$\omega$ by $E=\omega-I_p$, with $I_p$ being the ionization
potential of the target. Clearly the HHG signal $S(\omega)$ and
$W(E)$ depend on the laser properties. On the other hand,
$d(\omega)$ is the property of the target only. The factorization
in Eq.~(1) is most useful when one compares the HHG spectra from
two different targets in the identical laser field. Assuming that
the shape of $W(E)$ is species independent, by measuring the
relative HHG yields, one can deduce the PR cross section of one
species if the PR cross section of the other is known. The
validity of Eq.~(1) has been shown recently in Morishita {\it et
al.} \cite{toru} using HHG spectra calculated by solving the
time-dependent Schr\"odinger equation (TDSE) for atoms. The
validity of this factorization has also been shown for rare gas
atoms by Levesque {\it et al.} \cite{david} and for N$_2$ and
O$_2$ molecules \cite{hoang}, where the HHG spectra were
calculated using the SFA model. In the SFA the continuum electron
is approximated by plane waves, thus the dipole matrix elements
are calculated in the plane wave approximation (PWA).

In this Letter, we  have two goals. The first is to show that
electron wave packets obtained from the SFA model and from the
TDSE calculation are nearly identical, but the transition dipoles
calculated from PWA are significantly different from using
scattering waves (SW). This result suggests a scattering wave
based strong-field approximation (SW-SFA) for harmonic generation
where the wave packet is derived from the SFA but the transition
dipole is calculated using accurate SW. The second goal is to
check whether Eq.~(1) can be extended to the level of complex
amplitudes such that one can relate the phases of HHG to the
phases in the transition dipoles. Since phases of harmonics can be
measured \cite{salieres,midorikawa}, and they are needed in order
to incorporate the effect of propagation in the macroscopic
medium, such a study is important.

  First in Fig.~1 we compare the PR cross sections of Ar, Xe
  and Ne calculated by treating the continuum electrons using PWA
  to results calculated with accurate SW's.
  Clearly they show significant differences. They reflect
  the well-known facts that plane waves are poor approximations
  for representing continuum electrons in atoms and molecules for
  energies in the energy range of tens to hundreds eV's.

\begin{figure}
\mbox{\rotatebox{0}{\myscaleboxb{
\includegraphics{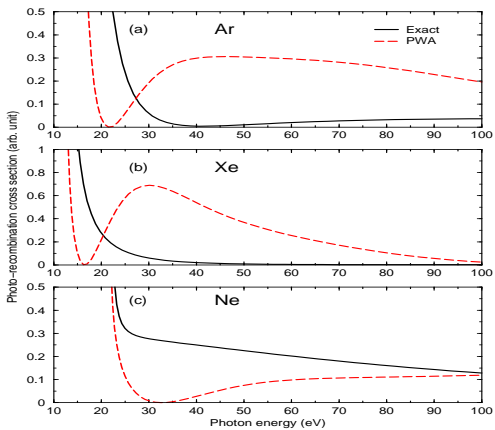}}}}
\caption{(Color online) Photo-recombination cross sections of Ar
(a), Xe (b) and Ne (c), obtained by using exact scattering
wavefunctions (solid curves) and within the plane-wave approximation
(dashed curves) for the continuum electrons.} \label{fig1}
\end{figure}

  To check whether the target structure affects the returning electron wave
  packet, in Fig.~2 we compare the wave packets $W(E)$ for Ne deduced from
  the TDSE and SFA results, using Eq.~(1). In the SFA case,
  the transition dipole is calculated within the PWA. Also shown is
  the $W(E)$ obtained from scaled atomic hydrogen, with the effective
  nuclear charge chosen such that the ionization potential of its
  1s ground state is the same as of Ne(2p).
  We used a laser pulse with duration (FWHM) of 10.3 fs,
peak intensity of $2\times 10^{14}$ W/cm$^2$ and mean wavelength
of 1064 nm. Note that we have normalized the results near the
cut-off. The normalization is to account for the difference in the
tunneling ionization rates from SFA and from TDSE, or from the
different species. This comparison shows that the shape, or the
energy dependence of the electron wave packets, depends only on
the laser parameters.

\begin{figure}
\mbox{\rotatebox{0}{\myscaleboxc{
\includegraphics{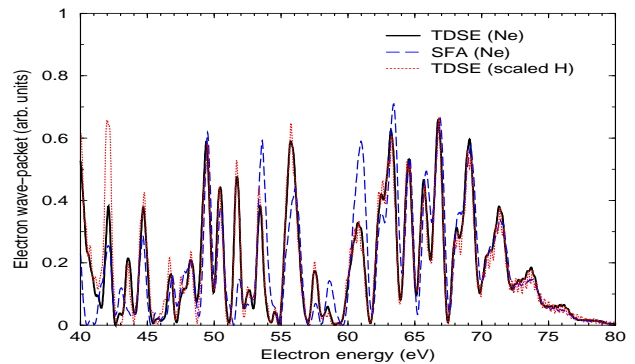}}}}
\caption{(Color online) Comparison of the returning electron
``wave packets'' extracted from the HHG spectra of Ne, obtained by
solving the TDSE (solid black line), and from the SFA model
(dashed blue line). Also shown is the TDSE result for the wave
packet from scaled H (dotted red line). For laser parameters, see
text.} \label{fig2}
\end{figure}

  Having established that the wave packet can be obtained from the
  SFA model, we now examine the accuracy of HHG calculated using the
  SW-SFA model where the wave packet is extracted from the SFA
  model and the transition dipoles are calculated using SW.
  In Fig.~3 we show the HHG spectra obtained from the TDSE, SFA,
and SW-SFA for Ar, Xe, and Ne. For Ar and Ne, the laser pulse has
peak intensity of $2\times 10^{14}$ W/cm$^2$ and mean wavelength
of 800 nm. The laser duration (FWHM) is 10 fs for Ar and 20 fs for
Ne. For Xe, the corresponding parameters are $5\times 10^{13}$
W/cm$^2$, 1600 nm, and 7.8 fs, respectively. The HHG yields for Ar
are shifted vertically in order to show their detailed structures.
For Ne and Xe, the SFA and SW-SFA results are normalized to the
TDSE results near the cutoff, i.e., close to $3.2U_p+I_p$, where
$U_p$ is the ponderomotive energy.

\begin{figure}
\mbox{\rotatebox{0}{\myscaleboxb{
\includegraphics{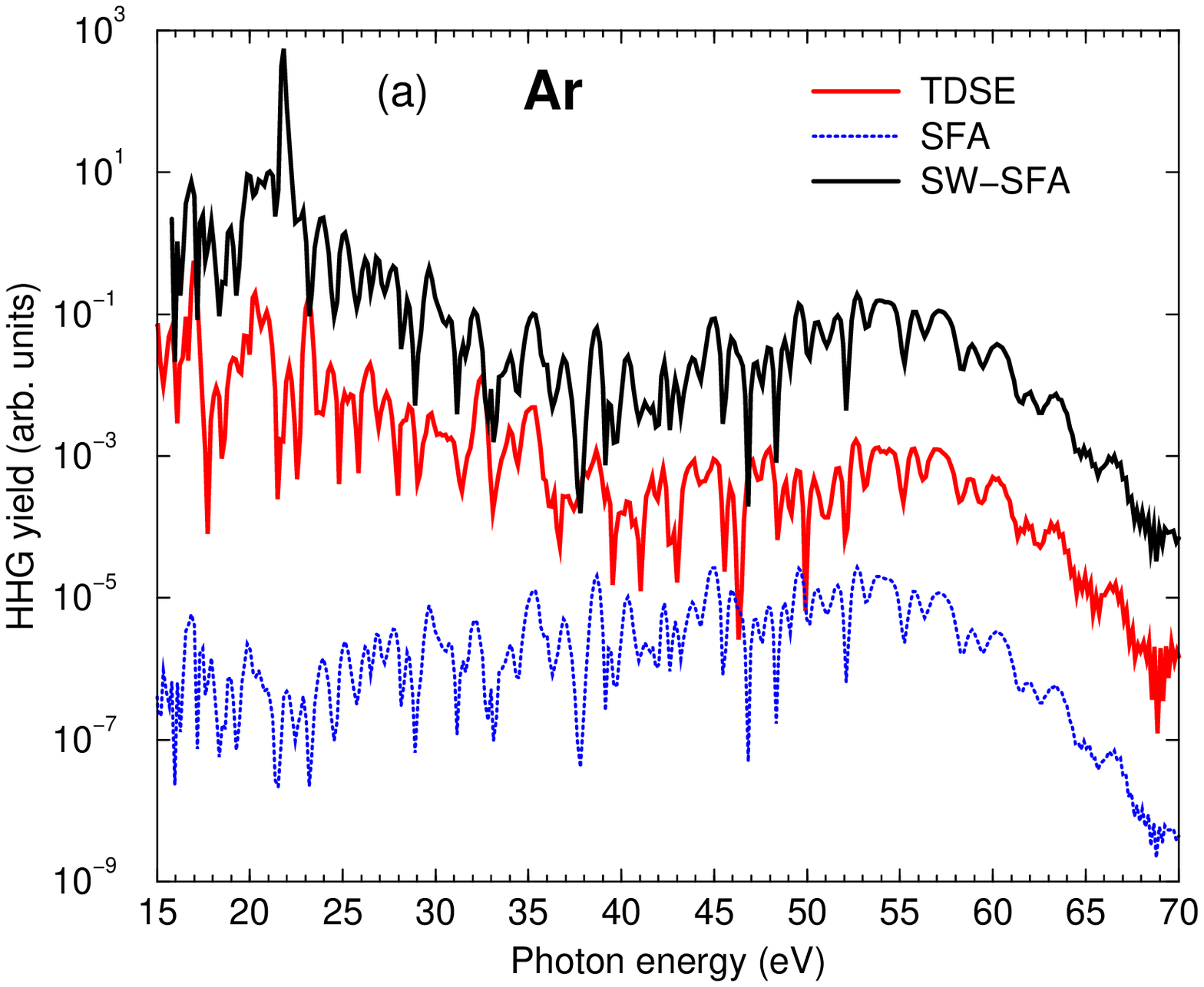}}}}
\mbox{\rotatebox{0}{\myscaleboxb{
\includegraphics{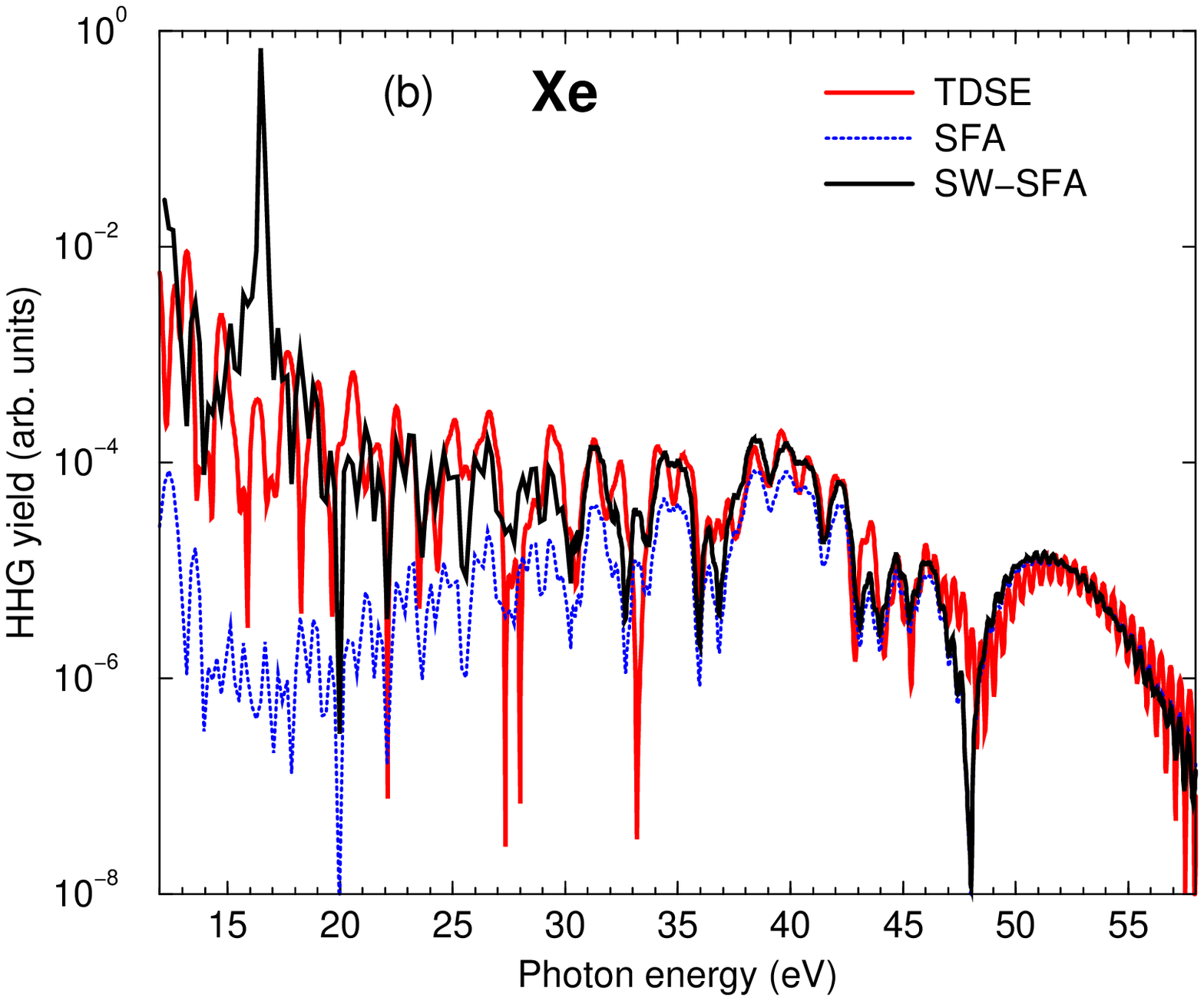}}}}
\mbox{\rotatebox{0}{\myscaleboxb{
\includegraphics{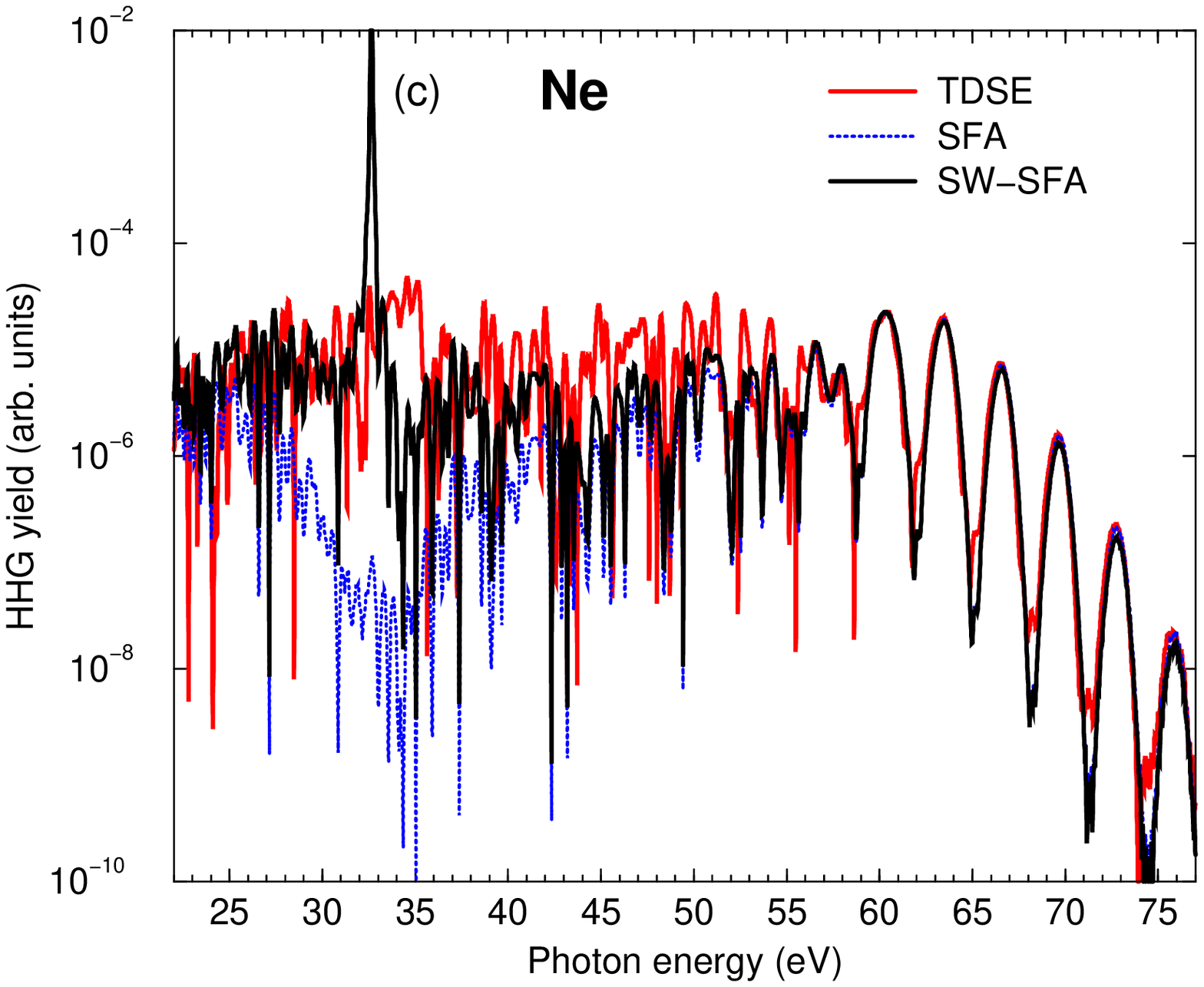}}}}
\caption{(Color online) Comparison of the HHG yields obtained from
numerical solution of the TDSE (solid red lines), the SFA (dashed
blue lines), and the SW-SFA model (solid black lines) for Ar (a),
Xe (b), and Ne (c). For laser parameters, see text.} \label{fig3}
\end{figure}

 The results in Fig.~3 clearly demonstrate the good improvement of
 the SW-SFA over the SFA in achieving better agreement with the TDSE
 results. Since the SFA gives the correct wave packet, its
 prediction would be ``reasonable" in the energy region where the
 dipole matrix element is rather flat, i.e., in the higher photon
 energy region. Thus the SFA would give adequate
 prediction of the HHG spectra usually near the cutoff region
 (after spectra are renormalized). This fact has been known \cite{lewenstein}.
 The improvement of SW-SFA occurs usually at lower photon
 energies where the PWA for the continuum
 electron is grossly incorrect. In particular, the transition
 dipole from PWA goes through zero at some lower energies, see
 Fig.~1. This is the energy region where the SFA suffers the largest
 errors. Because of the zeros in the dipole matrix elements in the
 PWA, the deduced wave packets from SFA would suffer large errors at
 the corresponding energies. These errors are reflected as the sharp
 spikes in the HHG spectra calculated using the SW-SFA model.

For a realistic description of the experimental harmonic spectra,
the effect of  phase matching and macroscopic propagation should
be addressed. To this end, the knowledge of the harmonic phase is
necessary. Thus extending Eq.~(1) to include the phase, can the
phase of the harmonic be expressed as the sum of the phase from
the wave packet and from the PR transition dipole? First we
establish that there is a close relationship between the harmonics
phase $\phi$ and the PR dipole phase $\delta$. To be specific, we
focus on Ar target.  We calculated the phase difference $\Delta
\phi$ for each harmonic generated from Ar and from its scaled
hydrogen (reference) partner under the same laser pulse. These
calculations were carried out using the TDSE with 4-cycle and
10-cycle laser pulses, intensities of $1$ and $2\times 10^{14}$
W/cm$^2$, and wavelength of $1064$ nm and $800$ nm. In Fig.~4(a)
we compare $\Delta \phi=\phi^{Ar}-\phi^{ref}$ with the PR dipole
phase difference $\Delta\delta=\delta^{Ar}-\delta^{ref}$. Here we
have shifted the harmonic phase difference to match the PR dipole
phase difference at $E=60$ eV. Clearly, the two agree very well
for the different lasers used. In particular, the phase jump near
40 eV (due to the Cooper minimum in Ar) is well reproduced. This
indicates that the phase of the wave packets from the two systems
are almost identical (up to a constant shift). Similar agreements
were also found for Xe and Ne, as shown in Fig.~4(b) and (c),
respectively. Here the laser pulses of 4 cycles duration are used,
other parameters are given as shown in the labels. This result
allows one to obtain the harmonic phase $\phi$ from the harmonic
phase of the partner atom $\phi^{ref}$ by using
$\phi=\phi^{ref}+\Delta\delta$.

\begin{figure}
\mbox{\rotatebox{0}{\myscaleboxb{
\includegraphics{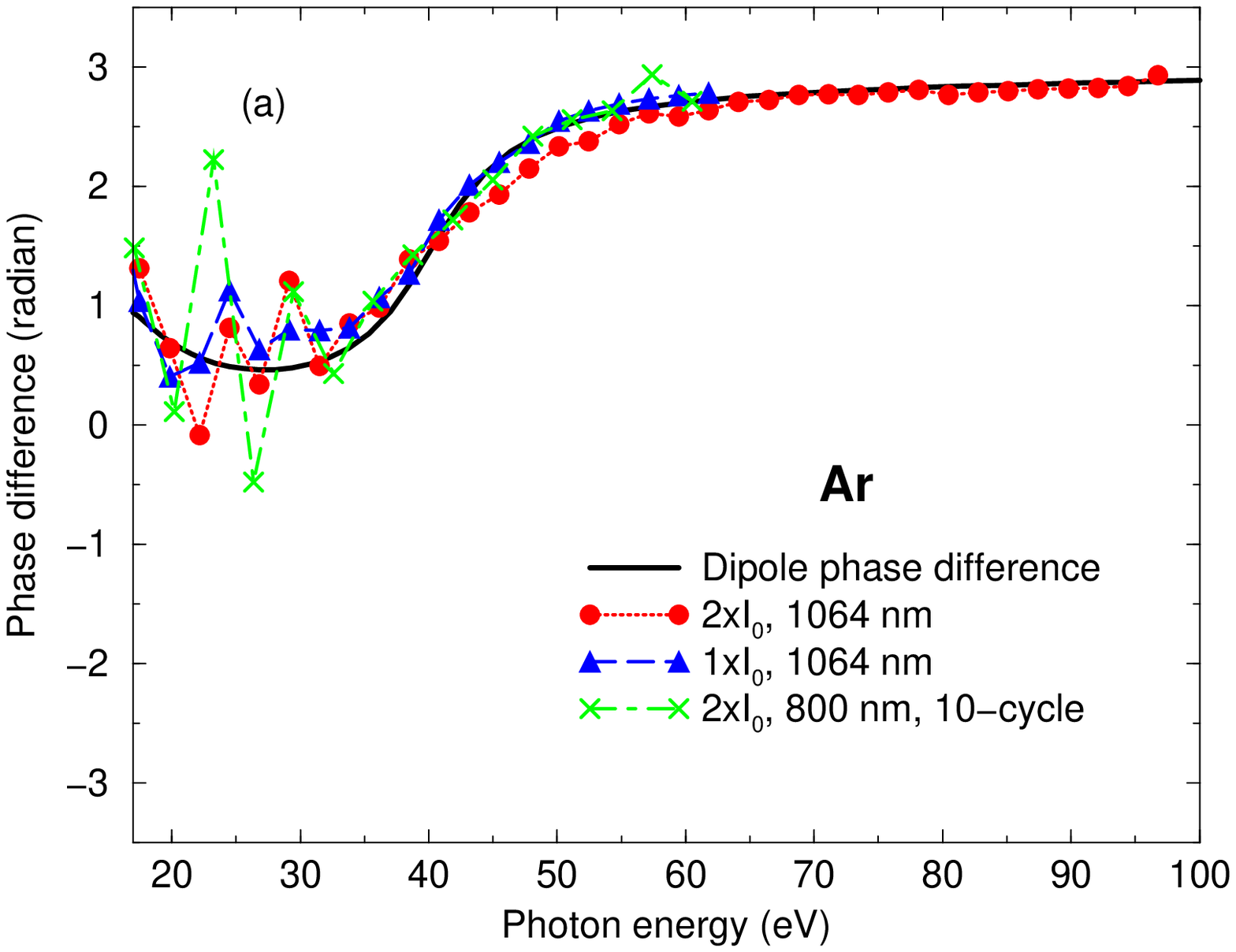}}}}
\mbox{\rotatebox{0}{\myscaleboxb{
\includegraphics{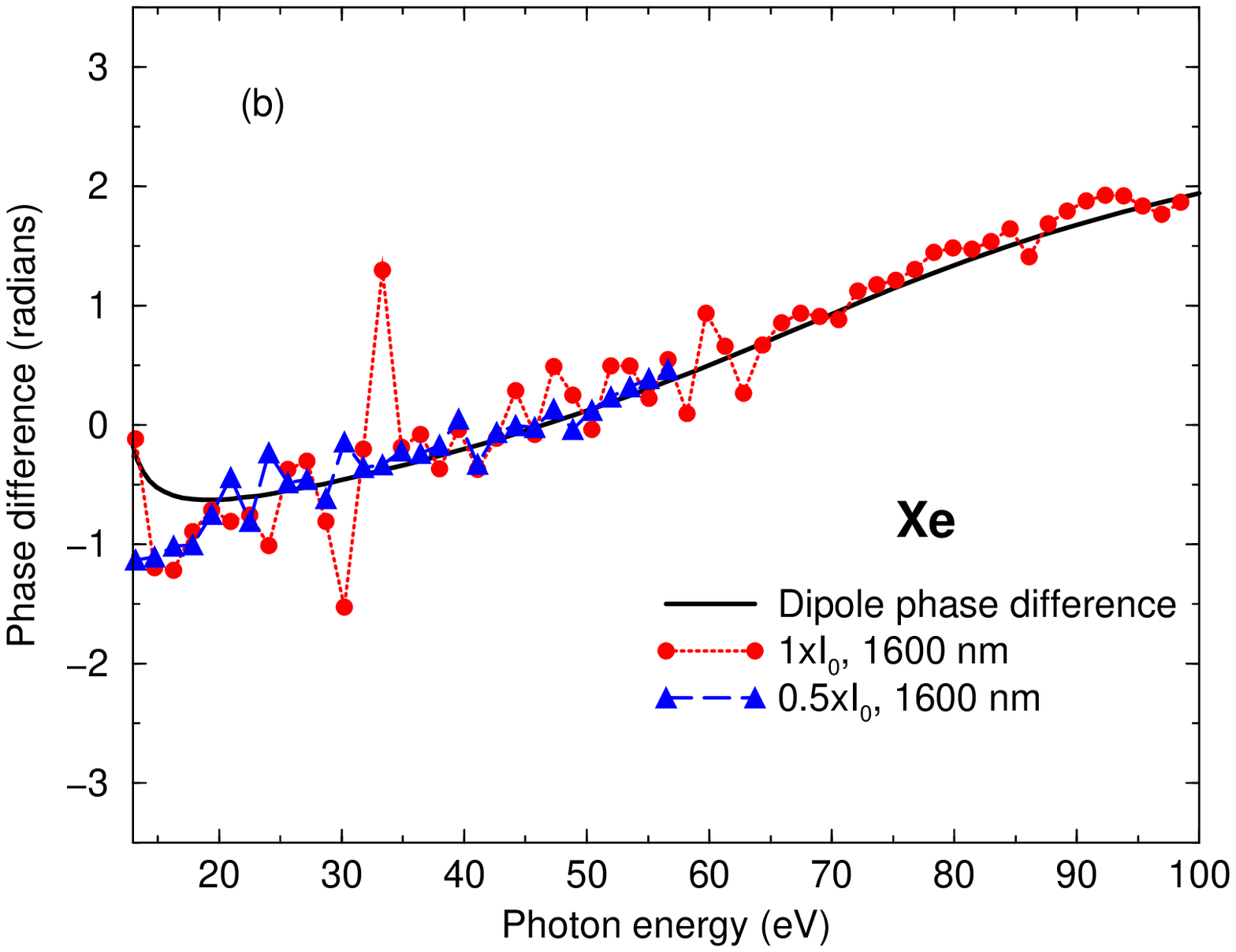}}}}
\mbox{\rotatebox{0}{\myscaleboxb{
\includegraphics{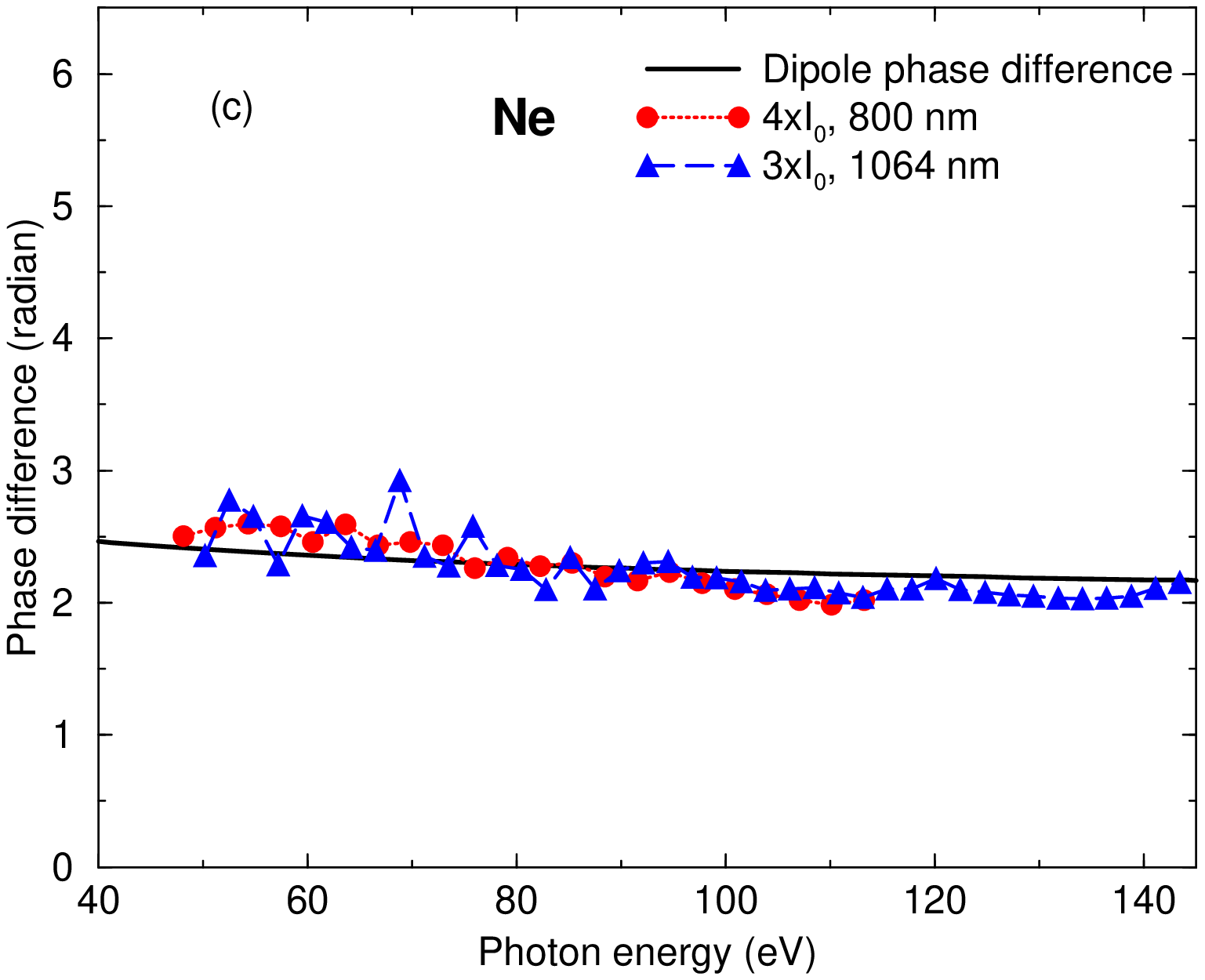}}}}
\caption{(Color online) Extracted harmonic phase difference
$\Delta \phi$ between Ar and scaled hydrogen obtained with
different lasers as function of emitted photon energy. The PR
dipole phase difference $\Delta\delta$ is given as solid black
line.(a) Ar, (b) Xe,and (c)Ne. I$_0$=10$^{14}$ W/cm$^2$.}
\label{fig4}
\end{figure}

 We have also applied the same procedure by comparing the TDSE
 and SFA results for the same target and found that
 $\Delta \tilde{\phi}=\phi^{TDSE}-\phi^{SFA}$ no longer agrees
 well with $\Delta \tilde{\delta}=\delta^{SW}-\delta^{PW}$.
 This indicates that the phase
 of the electron wave packet calculated from SFA differs from the
 one calculated by TDSE, even though their magnitudes agree well.
 How significant these differences affects the HHG spectra
 after macroscopic propagation? To this end we calculate the HHG
 spectra by coherently averaging
the induced polarization over an intensity range of the driving
laser. In Fig.~(5) we show the results for Ar from the TDSE, the
SW-SFA, and the one with the wave packet extracted from the scaled
hydrogen. All of these results are coherently averaged over 11
equally-spaced intensities in the range from 1.8 to $2.2\times
10^{14}$ W/cm$^2$. The laser is of 800 nm wavelength and 30 fs
(FWHM). The scaled H result is indeed in quite good agreement with
the exact TDSE calculations. This is not surprising since we have
shown that the phase of the wave packet from the scaled H and from
Ar are almost identical at a single intensity. For SW-SFA, the
agreement is not as good, but the improvement over SFA is still
significant. The phase in SFA (or SW-SFA) can probably be improved
by adding some correction to the semi-classical action, for
example, as has been suggested \cite{ivanov,smirnova}. At present,
it is better to extract the phase of the wave packet from the
companion atomic target where TDSE calculations can be carried
out.

\begin{figure}
\mbox{\rotatebox{0}{\myscaleboxc{
\includegraphics{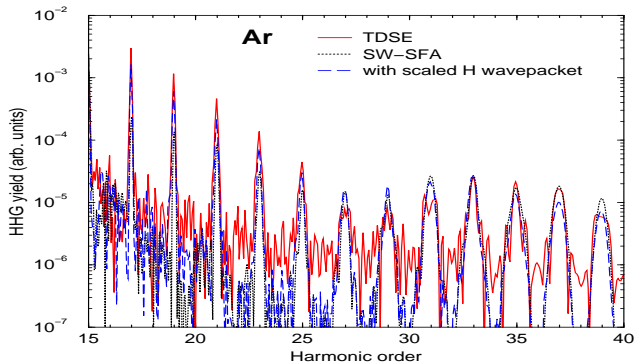}}}}
\caption{(Color online) HHG spectra for Ar from the ``simulated''
macroscopic propagation. Shown are results from the exact TDSE
(solid red line), SW-SFA(dotted black), and by using the wave
packet extracted from TDSE solution for scaled H(1s) (dashed
blue). For laser parameters, see text.} \label{fig5}
\end{figure}

 Here we comment on the computational details. The solution of the TDSE and
 the choice of one-electron model potential for describing the
 atom have been described previously \cite{chen,toru1}.
 The electric field of the laser pulse is written
 in the form  $E(t)=E_0a(t)\cos(\omega t)$,
with the envelope given by $a(t)=\cos^2(\pi t/\tau)$, where $\tau$
is 2.75 times the FWHM of the laser pulse. To calculate the PR
cross section, the scattering wavefunction is expanded in terms of
partial waves and the transition dipole is calculated for the
continuum electron that has the wave vector along the polarization
axis only.

Before concluding we mention several earlier related works. There
exists a wealth of literature aiming at improving the SFA model,
e.g., by including Coulomb distortion \cite{ivanov,kaminski96}, or
by eikonal approximations \cite{smirnova}. In these approaches,
the PR processes are still treated approximately. For example, use
of Coulomb wave for the continuum electron would not produce the
Cooper minimum in the PR cross section in Ar (see Fig.~1). The
advantage of SW-SFA is that it factors out the target structure
explicitly. A minimum in the HHG spectra may be attributed to the
minimum in the PR cross section and this position should not
change with laser parameters. Such minima are of particular
interest for molecular targets since minima in the molecular
dipole matrix element may be interpreted as due to the
interference between the emission amplitudes from different atomic
centers. Interference minima have been observed experimentally in
CO$_2$ by different groups \cite{kanai,vozzi}, but the observed
positions of the minimum are not identical and thus other possible
interpretations have been suggested \cite{atle}. Another hot
topics in recent years is the tomographic method for imaging the
molecular orbitals \cite{itatani}. This pioneering work deduced
the dipole matrix elements of N$_2$ molecules by comparing the HHG
spectra of N$_2$ vs Ar, using the factorization Eq.~(1). In order
to use the tomographic procedure to obtain the ground state
wavefunction of N$_2$, they approximated the continuum
wavefunctions of Ar and N$_2$ by plane waves. In view of Fig.~1,
their success of extracting good valence orbital wavefunction of
N$_2$ is surprising. We note, however, in Itatani {\it et al}
\cite{itatani}, the continuum electron energy is set equal to the
photon energy, arguing that the electron recombining near the core
should gain the additional binding energy. For Ar, this would
shift the PWA curve in Fig.~1 by 15.7 eV, making the PWA result
much closer to the SW result. However, this shift does not always
work, see Xe and Ne examples in Fig.~1.

 In conclusion, we have established quantitatively that the last step of the
 three-step model of HHG can indeed be
 expressed as the photo-recombination (PR) process of
 the returning electron wave packet. The wave packet depends
 nonlinearly on the laser, but its shape and phase are largely
 independent of the target. Thus if the PR of a reference target is known, the PR of
 another target can be derived by measuring the HHG of the two
 species under identical laser pulses. Since the results should be
 independent of the lasers, this allows for an important check on
 the accuracy of the measurements. We also showed that the HHG
 spectra can be calculated using the SW-SFA model. This model
 describes well the single atom HHG intensity, but the phase
 needs further corrections. For complex systems, SW-SFA would be a
 good starting point for describing the HHG spectra since the
 PR process is accurately incorporated. While our conclusion has
 been derived based on atomic targets and in the single active
 electron model, we anticipate that the results
 are applicable to molecules where accurate TDSE calculations are not available in general.
 The present result offers a systematic roadmap for extracting target structure information from
the high-order harmonics generated by intense lasers.

This work was supported in part by the Chemical Sciences,
Geosciences and Biosciences Division, Office of Basic Energy
Sciences, Office of Science, U. S. Department of Energy.
TM is also supported by a Grant-in-Aid for Scientific Research (C) from
MEXT, Japan,  by the 21st Century COE program on ``Coherent
Optical Science'',  and by a JSPS Bilateral joint program between
US and Japan.

\end{document}